\documentclass[a4paper,12pt]{article}
\usepackage[left=2cm,right=2cm,
top=2cm,bottom=2cm]{geometry}
\usepackage{graphicx}
\usepackage{floatrow}
\usepackage{latexsym}
\usepackage{amssymb,latexsym,amsmath}
\usepackage{latexsym}
\usepackage{xcolor}
\usepackage{hyperref}
\usepackage{caption} 
\usepackage[export]{adjustbox}
\usepackage{color}
\usepackage{colortbl}
\usepackage{authblk}
\usepackage{etoolbox}
\usepackage{caption}
\usepackage{floatrow}
\usepackage{cite}
\DeclareCaptionFormat{simple}{#1#2#3\hrulefill}
\begin{document}
	
	\title{Low-frequency vibrations of water molecules in minor groove of the DNA double helix
	}
	
	\author{T.L. Bubon      and
		S.M. Perepelytsya
	}
	\affil{Bogolyubov Institute for Theoretical Physics of the National Academy of Sciences of Ukraine, 14-b Metrolohichna Str., Kyiv 03143, Ukraine}
	\date {\today}
	\maketitle
	
	\begin{abstract}
		The dynamics of the structured water molecules in the hydration shell of the DNA double helix is of paramount importance for the understanding of many biological mechanisms. In particular, the vibrational dynamics of a water spine that is formed in the DNA minor groove is the aim of the present study. Within the framework of the developed phenomenological model, based on the approach of DNA conformational vibrations, the modes of H-bonds stretching, backbone vibrations, and water translational vibrations have been established. The calculated frequencies of translation vibrations of water molecules vary from 167 to 205 cm$^{-1}$ depending on the nucleotide sequence. The mode of water vibrations higher than the modes of internal conformational vibrations of DNA. The calculated frequencies of water vibrations have shown a sufficient agreement with the experimental low-frequency vibrational spectra of DNA. The obtained modes of water vibrations are observed in the same region of the vibrational spectra of DNA as translation vibrations of water molecules in the bulk phase. To distinguish the vibrations of water molecules in the DNA minor groove from those in the bulk water, the dynamics of DNA with heavy water was also considered. The results have shown that in the case of heavy water the frequencies of vibrations decrease for about 10 cm$^{-1}$ that may be used in the experiment to identify the mode of water vibrations in the spine of hydration in DNA minor groove.	
	\end{abstract}
	
	\section{Introduction}
	\label{sec:1}
	The natural DNA consists of two chains of nucleotides (adenine, guanine, thymine, and cytosine) winding around each other as the double helix \cite{saengerprinciples}. The hydrophobic nucleotide bases form the complementary H-bonded pairs (A-T and G-C) inside the macromolecule to reduce the contact with water molecules, while the negatively charged phosphate groups of the double helix backbone are exposed to the solution to be neutralized by the positively charged ions of metals. Starting with the very first X-ray and modeling studies of the DNA molecular structure, the ion-hydration environment was known to stabilize the macromolecule structure  \cite{watson1953molecular,franklin1953structure,wilkins1963molecular}. The hydration shell of DNA macromolecule ranges over several layers from the surface within the region of 10--15 {\AA} and can be conditionally divided into primary and secondary hydration shells \cite{blagoi1991metal,via1993physical,manning1998counterion}. The conformational dynamics of the DNA double helix and the dynamics of its hydration shell are interrelated \cite{hynes2016dynamical,yonetani2012determines} that is determinative for the mechanisms of nucleic-protein recognition and the interaction with the biologically active compounds  \cite{jayaram2004role,haq2002thermodynamics,billeter1996hydration}. Thus, to understand the physical mechanisms of DNA biological functioning the structure and dynamics of the hydration shell of the double helix should be studied.
	
	Under the physiological conditions, DNA takes the specific \emph{B}-form with a hydration shell consisted of about 30--50 water molecules per nucleotide pair \cite{via1993physical,tao1989structure,lavalle1990counterion}. The hydration shell of DNA macromolecule has different structure in different compartments of the double helix: minor groove, major groove, and outer region of the macromolecule. Earliest X-ray studies of the structure of the crystals of  DNA fragment d(CGCGAATTCGCG), known as Drew-Dickerson dodecamer, showed that in AATT nucleotide region in the minor groove the water molecules are ordered in the specific structure known as spine of hydration \cite{drew1981structure}. The hydration spine is composed of water molecules bridging N$_{3}$ and O$_{2}$ atoms of purines (A, G) and pyrimidines (C, T) nucleotide bases of opposite strands of DNA. Such structure has been also observed in further X-ray experiments \cite{tereshko1999hydrat}, nuclear magnetic resonance (NMR) studies \cite{kubinec1992nmr,liepinsh1992nmr} and chiral nonlinear vibrational spectroscopy \cite{mcdermott2017dna}. The formation of the hydration spine was also observed in molecular dynamics (MD) simulation studies \cite{chuprina1991molecular,duan1997molecular}. 
	
	The dynamics of hydration shell is heterogeneous and depends on water molecule location in the double helix. The residence time of water molecules near phosphate groups are about 10 ps \cite{siebert2015anharmonic,floisand2015computational}, while in the bulk water it is about 1 ps \cite{jimenez1994femtosecond,asbury2004water}. In the grooves of the double helix the dynamics of water molecules is even slower than near phosphate groups. In the minor groove, the residence time of water molecules is very long (more than 1 ns) compared to those in the major groove (0,5 ns) \cite{liepinsh1992nmr}. However, the NMR studies  \cite{denisov1997kinetics,phan1999determination}  showed that the residence time of water molecules in the minor groove is within the range  0.2--0.9 ns, while in the major groove it is about 0.1 ns. These results quantitatively agree with MD simulations, where very slow water molecules in the minor groove (residence time more than 100 ps) and in the major groove (about 60 ps) \cite{hynes2016dynamical,saha2015distribution} were detected. The residence time of water molecules is determined by the lifetime of H-bonds that are formed with the atoms of DNA. The MD simulation studies showed that in the minor groove the lifetime of water H-bonds is about 50 ps, while in the major groove it is more than two times lower \cite{yonetani2012determines,pal2006exploring}. The characteristic residence times of water molecules in the DNA binding sites are much longer than the period of water vibrations ($<$ 1 ps). Therefore, the vibrations of water molecules as whole around DNA are expected to be observed in the vibrational spectra. 
	
	The translation vibrations, i.e H-bonds stretching, of water molecule and H-bonds biding are observed in the spectra region 60--200 cm$^{-1}$, while librational (hindered rotational) motions are above 300 cm$^{-1}$ \cite{walrafen1964raman,walrafen1996low,ohmine1999water,galvin2011extreme}. In the Raman spectra, two bands at about 175 and 60 cm$^{-1}$ are observed that attributed to H-bonds stretching and bending vibration, respectively \cite{walrafen1964raman,walrafen1990raman,walrafen1996low}. These vibrations are also observed in the infra-red absorption (IR) spectra \cite{draegert1966far,vij2004far,marechal2011molecular}, neutron scattering experiments \cite{safford1969investigation}, analytical estimations \cite{gaiduk2001concept} and MD simulations \cite{torii2011intermolecular,heyden2010dissecting}. At the same spectra range the DNA conformational vibrations characterized by the displacements of the atomic groups in the nucleotide pairs are observed \cite{lamba1989low,weidlich1990raman}. The DNA low-frequency spectra are characterized by the narrow peak about 20 cm$^{-1}$ and a broad band near 85 cm$^{-1}$ \cite{urabe1981low,urabe1985collective,weidlich1990raman}. In a higher frequency range (100--200 cm$^{-1}$) the modes that depend on concentration and type of the counterions have been observed \cite{powell1987investigation,weidlich1990counterion}. The low-frequency spectra of the DNA have been described by different analytical approaches \cite{young1989calculation,volkov1987conformation,volkov1991theory,matsumoto1999dynamic,cocco2000theoretical}. In particular, the detailed interpretation of the DNA low-frequency spectra was done within the framework of the approach of the conformational vibrations of the double helix \cite{volkov1987conformation,volkov1991theory}. According to this approach, the lowest mode around 20 cm$^{-1}$ is related to the vibrations of DNA backbone, while the broad band near 85 cm$^{-1}$ characterizes the vibrations of H-bond stretching in the bases pairs and intranucleoside vibrations due to the conformational flexibility of sugar rings. The ion-depended modes in DNA low-frequency spectra \cite{powell1987investigation,weidlich1988counterion,weidlich1990counterion} have been determined as shown to be related to the vibrations of counterions with respect to the phosphate groups of the double helix backbone (ion-phosphate vibrations) \cite{perepelytsya2004ion,perepelytsya2007counterion,perepelytsya2010low}. To determine the modes of water molecules vibrations in the DNA low-frequency spectra, the theoretical approach should be developed. 

The goal of the present study was to determine the low-frequency vibrations of water molecules in the DNA minor groove. To solve this problem, the analytical model has been elaborated based on the phenomenological approach of the DNA conformational vibrations \cite{volkov1987conformation}. The developed model is described in the Section \ref{sec:2}. In the Section \ref{sec:3}, the parameters of the model have been determined. In the Section \ref{sec:4}, the frequencies and amplitudes of vibration of DNA with water molecules have been calculated. As a result, the mode of water translational vibrations in the DNA minor groove has been established within the frequency range 167--205 \ cm$^{-1}$ depending on nucleotide sequence. In the Section \ref{sec:5}, the obtained frequencies of vibrations have been compared with the experimental spectra of DNA in an aqueous environment. The frequencies of water vibrations in the DNA minor groove in the same spectral range as translation vibrations of bulk water. To distinguish the vibrations of water molecules in the minor groove from those in the bulk, the vibrations of DNA with heavy water have been considered.
	
	\begin{figure} 
		\includegraphics[width=0.75\textwidth]{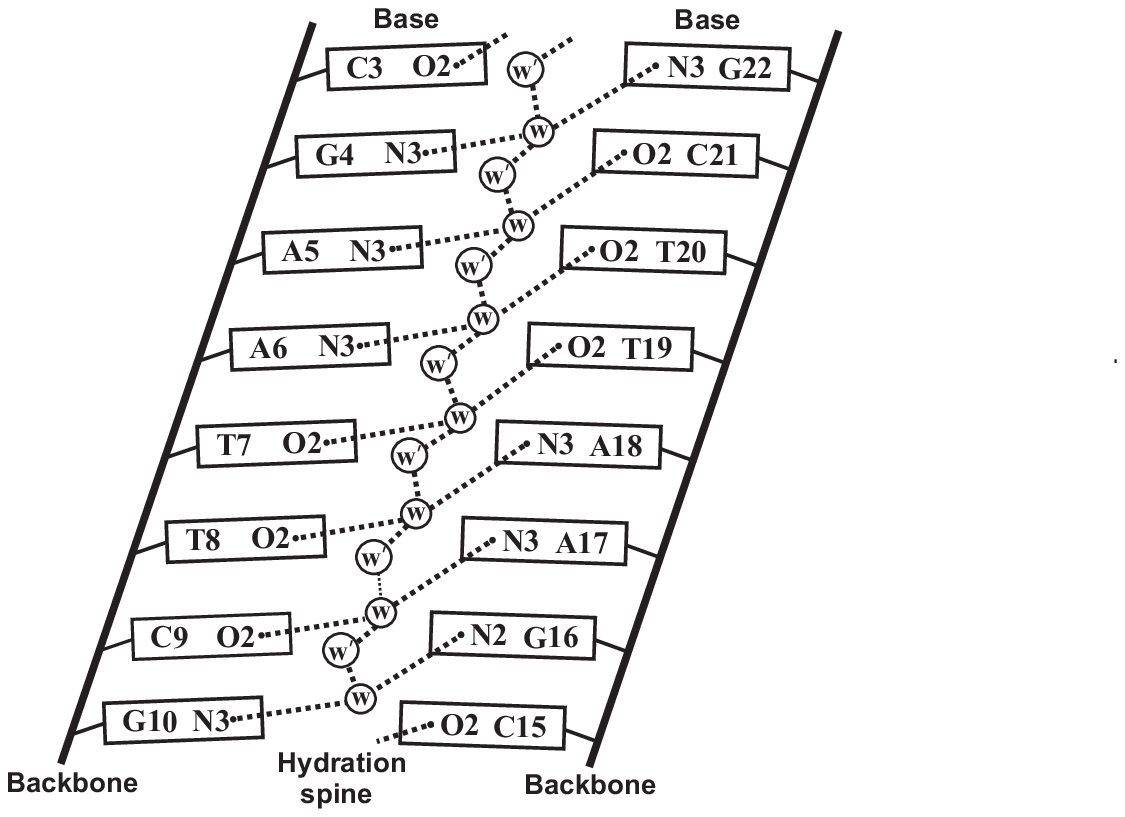}
		\caption{\small{The scheme of the structure of the first and second layers of water molecules in the hydration spine in the DNA minor groove according to \cite{drew1981structure,tereshko1999hydrat}. Circles with w and w' indicate water molecules of the first and second layer, respectively. The atoms of acceptors of the nucleotide bases are indicated.}}
		\label{fig:1}
	\end{figure}
	
	\section{Model of  water  vibrations in  DNA minor groove}\label{sec:2}
\begin{figure}
		\includegraphics[width=0.85\textwidth]{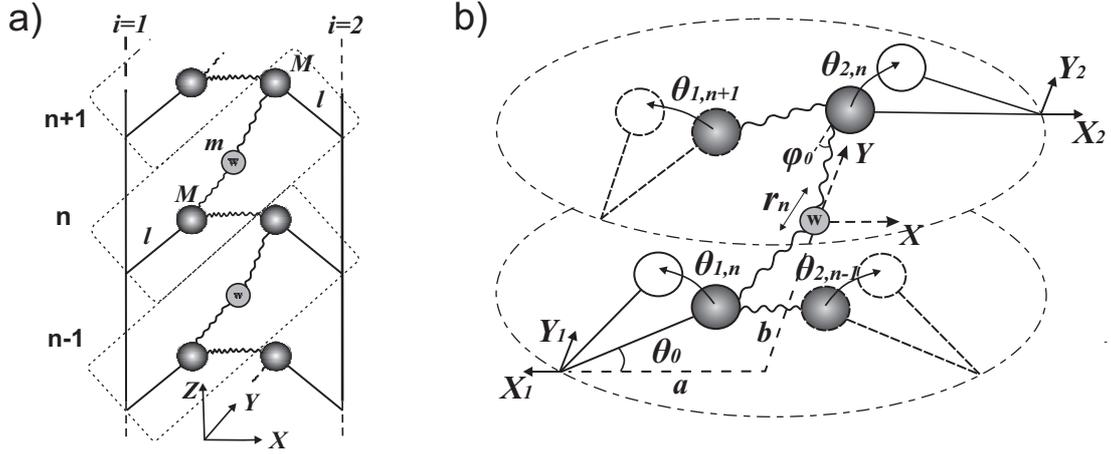}
		\caption{\small{Model for the vibrations of water molecules in DNA minor groove. a) The chain of monomer links consisted of two nucleotides bounded by one water molecule. The monomer links are shown by dotted frames. $\ell$ is the reduced length of physical pendulum; $M$ and $m$ are masses of nucleoside and water molecule, respectively; $i$ and $n$ enumerate the monomer links and chains of the double helix, respectively. b) Two pairs of complementary bases in the plane orthogonal to the axis of the double helix. The nucleotides bridged by water molecule represent the selected monomer link (water-bounded pair). Water molecule placed in the centre of the coordinate system. ${\phi}_{0}$ is the equilibrium angle between nucleoside and water molecule masses. ${\theta}_{0}$ is the equilibrium angle between nucleoside mass and backbone; ${\theta_{i,n}}$ and $r_{n}$ are vibrational coordinates of the model. The arrows indicate positive directions of displacements. $a$ and $b$ are distances from the centre of coordinate system to suspension point (backbone) of nucleosides mass on $OX$ and $OY$ direction, respectively.}}
		\label{fig:2}
	\end{figure}

	To built the model of vibrations of water molecules in the hydration spine of DNA minor groove the approach for the description of the conformation dynamics of DNA double-helix developed by Volkov and Kosevich \cite{volkov1987conformation} has been used. In the present model, the sugar-phosphate backbone of the double helix is modelled by rigid walls. The nucleosides with mass $M$ are presented as the physical pendulums with the reduced length $\ell$ suspended to the backbone at the angle $\theta_{0}$. The physical pendulums rotate in the plane orthogonal to the helical axis. The nucleoside bases in different DNA stands are connected by the hydrogen bonds modelling the nucleotide pairs. Water molecules are represented as the masses bonded to the pendulum-nucleosides, bridging the nucleotide bases of one stand with the nucleotide base of another stand (Fig.\ref{fig:1}). The nucleosides of opposite stands bridged by water molecule form the monomer link in our model (Fig.\ref{fig:2}a). The nucleic bases that interact with the water molecules belong to different nucleotide pairs will be referred as water-bounded pairs. The interactions between complementary bases in the nucleotides pairs and interaction of the bases among chain are taken into account. In the present work, the nucleotides of different types (adenine, thymine, guanine and cytosine) have been characterized by averaged values of the parameters $\theta_0,\ell$ and $M$.
	
 To consider the dynamics of structural elements of presented model, the following system of coordinates has been used. The centre of coordinates is related to the water molecule. The axis $OZ$ is directed along the axis of the double helix. The axis $OY$ is directed to the major groove. The axis $OX$ is orthogonal to the plane $YOZ$. The nucleotides pairs placed in the plane $XOY$, where axis $OX$ defines the positive direction for each nucleoside to the outside of the double helix. The vibrations of nucleosides with respect to the phosphate groups are described by deviations $\theta_{i,n}$ from the equilibrium angle $\theta_0$, where $n$ enumerates the monomer link $(n= 0,\pm 1,..,\pm N)$ and $i$ is the number of DNA strand ($i = 1,2$). The position of nucleosides with respect to the centre of coordinate system are characterized by the equilibrium distances $a$ and $b$ in $OX$ and $OY$ directions, respectively. The position of water molecules with mass $m$ with respect to nucleotides is characterized by the equilibrium angle $\phi_0$ describing the orientation of water molecule. The displacements of water molecules from the equilibrium positions in the monomer link are described by the deviation $r_n$ (Fig.\ref{fig:2}b). The coordinates of the displacements for the system are presented as follows:	
	\begin{equation}\label{eq1}
	X_{n,i} =a-\ell\cos(\theta_{0}+\theta_{i,n});
	\end{equation}
	\begin{equation}\label{eq2}
	Y_{n,i}=\ell\sin(\theta_{0}+\theta_{i,n})-b;
	\end{equation}
	\begin{equation}\label{eq3}
	u_{x,n}=r_{n}\cos\phi_{0};
	\end{equation}
	\begin{equation}\label{eq4}
	u_{y,n}=r_{n}\sin\phi_{0}.
	\end{equation}

	Within the framework of the introduced coordinates, the energy of vibrations of structural elements of the double helix and water molecules may be written as follows:
	\begin{equation}\label{5}
	E = \sum_{n}\left[K_{n}+U_{n}+U_{n-1}\right],
	\end{equation}
	where $K_{n}$ and $U_{n}$  are the kinetic and potential energies of the monomer link $n$; $U_{n-1}$ describes the interactions in the complementary pairs and interactions of the staked bases along the chain. 
	
	Taking into account that in the present work  small displacements of the masses in the model are considered the coordinates (\ref{eq1})-(\ref{eq4}), the kinetic energy may be written in the following form:
	\begin{equation}\label{eq6}
	K_{n}=\frac{1}{2} \left[ \sum_{i=1}^{2} M(\ell \dot{\theta}_{i,n})^{2}+m\dot{r}_{n}^2\right].
	\end{equation}
	
	The potential energy of the monomer link in harmonic approximation may be written as follows:
	\begin{equation}\label{eq7}
	\begin{split}
	U_{n}=\frac{1}{2}\sum_{i=1}^{2} \Bigl[\beta\theta_{i,n}^2+k\left(\theta_{i,n}\ell C_{1}+r_{n}C_{2}\right)^2\Bigr].
	\end{split}
	\end{equation}
	Here 
	$C_{1}=\left(a\sin\theta_{0}-b\cos\theta_{0})\right /R_{0}$ and   $C_{2}=\left(a\cos\phi_{0}+b\sin\phi_{0}-\ell \cos(\theta_{0}-\phi_{0})\right)/R_{0}$, where $R_{0}$ is the equlibrium distance in water-bounded pair and  $R_{0}=\left(a^2+b^2+\ell^2-2(a\ell_{b}+b\ell_{a})\right)^{1/2}$. $\ell_{a}=\ell \sin\theta_{0}$ and $\ell_{b}=\ell \cos\theta_{0}$. $\beta$ is the force constant of the rotation of nucleosides with respect to the backbone chain in the plane of nucleosides pair without interaction between adjacent bases; $k$ is the force constant of changes in the equilibrium distance between water molecule and nucleoside.
	
	The potential energy of displacements in complementary nucleotide pairs and the energy of interaction along the chain may be written as follows:
	\begin{equation}\label{eq8}
	U_{n-1}=\frac{1}{2}\Bigl[\alpha \ell_{a}^2\left(\theta_{1,n}+\theta_{2,n-1}\right)^2+\sum_{i=1}^{2}g\left(\theta_{i,n}-\theta_{i,n-1}\right)^2\Bigr],
	\end{equation}
	where $\alpha$ is the force constant of H-bonds stretching in complementary pairs; $g$ is the force constant of interactions between the stacked adjacent base pairs.
	
	Using the formulae (\ref{eq6})-(\ref{eq8}), the equations of motion have been obtained:
	\begin{equation}\label{eq9}
	\begin {cases}
	
	M\ell^2\ddot{\theta}_{1,n}+\beta\theta_{1,n}+\alpha \ell_{a}^2(\theta_{1,n}+\theta_{2,n-1})+\\
	+g(\theta_{1,n-1}-2\theta_{1,n}+\theta_{1,n+1})
	+k\ell C_{1}(\theta_{1,n}\ell C_{1}+r_{n} C_{2})= 0;\\
	M\ell^2 \ddot{\theta}_{2,n}+\beta\theta_{2,n}+\alpha \ell_{a}^2(\theta_{1,n-1}+\theta_{2,n})+\\+g(\theta_{2,n-1}-2\theta_{2,n}+\theta_{2,n+1})+k\ell C_{1}(\theta_{2,n}\ell C_{1}+r_{n} C_{2})= 0;\\
	m\ddot{r}_{n}+kC_{2}(\theta_{1,n}\ell C_{1}+r_{n}C_{2})+kC_{2}(\theta_{2,n}\ell C_{1}+r_{n}C_{2})=0.
	
	\end{cases}
	\end{equation}
	
The system of equations (\ref{eq9}) consists of the second-order differential equations. The solutions of the equations (\ref{eq9}) may be found in the following form:
	\begin{equation}\label{eq10}
	\theta_{i,n}= \tilde{\theta}_{i}e^{i(\omega t-\varkappa n)},
	r_{n}=\tilde{r}e^{i(\omega t-\varkappa n)},
	\end{equation}
	where $\varkappa$ is the projection of the wave vector to the axis $Z$; $\tilde{\theta}$ and $\tilde{r} $ are the amplitudes of vibrations.
	
In the present work, the modes observed in the experimental vibrational spectra in the scope of interest. Therefore, we interested in finding the vibrations of optic type. In the case of the approach of dynamical lattice theory, this is equivalent to the long-wave limit ($\varkappa\to 0$). Using the long-wave approximation, the equations (\ref{eq9}) in new variables $\tilde{\theta}_{1}+\tilde{\theta}_{2} =\tilde{\theta}$ and $\tilde{\theta}_{1}-\tilde{\theta}_{2}=\tilde{\xi}$ are split as follows:
	\begin{equation} \label{eq11}
	\begin{split}
	\begin {cases}
	\tilde{\theta}l\left(\omega^2-\beta_{0}-k_{0}C_{1}^2\frac{m}{M}-2\alpha_0 \sin^2 \theta_0\right)+ \tilde{r}\left(-2k_{0}C_{1}C_{2}\frac{m}{M}\right)=0;\\
	\tilde{\theta}\ell \left(-k_{0}C_{1}C_{2}\right)+\tilde{r}\left(\omega^2-2k_{0}C_{2}^2\right)=0.
	\end{cases}
	\end{split}
	\end{equation}
	\begin{equation} \label{eq12}
	\tilde{\xi}\left(\omega^2-\beta_{0}-k_{0}C_{1}^2\frac{m}{M}\right)=0,
	\end{equation}
		where $ \alpha_{0}=\alpha/M, \beta_{0}=\beta/M\ell^2, k_{0}=k/m.$ 
	
Using the existence condition for a solution of equations (\ref{eq11}), the expression for frequencies of longwave vibrations has been determined in the form:
	\begin{equation}\label{eq13}
	\begin{split}
	\omega^4-\omega^2&\Bigl(\beta_0+2\alpha\sin^2\theta_{0}+k_{0}
	\bigl(2C_{2}^2+C_{1}^2\frac{m}{M}\bigr)\Bigr)+\\
	&+2k_{0}C_{2}^2\bigl(\beta_{0}+2\alpha_{0}\sin^2\theta_{0}\bigr)=0.
	\end{split}
	\end{equation}
	
	From the equation (\ref{eq12}) the following equation for frequency has been determined:
	\begin{equation}\label{eq14}
	\omega^2-\beta_{0}-k_{0}C_{1}^2\frac{m}{M} =0.
	\end{equation}
	
The solutions of the equation (\ref{eq13}) have been found in the following form:
	\begin{equation}\label{eq15}
	\begin{split}
	\omega_{1,2}^2=
	\frac{1}{2}
	\Bigl\{
	\beta_{0}+2\alpha_{0}\sin^2\theta_{0}+k_{0}(C_{1}^2\frac{M}{m}+2C_{2}^2)\pm\\
	\qquad
	\pm
	\bigl[
	\bigl(
	\beta_{0}+2\alpha_{0}\sin^2\theta_{0}+k_{0}
	\bigl(
	C_{1}^2\frac{M}{m}+2C_{2}^2
	\bigr)
	\bigr)^{2}\\
	-8k_{0}C_{2}^2(\beta_{0}+2\alpha_{0}\sin^2\theta_{0})
	\bigr]^{1/2}
	\Bigr\}.
	\end{split}
	\end{equation}
	
	The solution of equation (\ref{eq14}) has been obtained as follows:
	\begin{equation}\label{eq16}
	\omega_{3}^2=\beta_{0}+k_{0}C_{1}^2\frac{m}{M}.
	\end{equation}
	
	According to the character of frequency of vibration of the DNA with water molecules, the vibrational modes $\omega_1,\omega_2,\omega_3$ can be classified as the modes of water molecules vibrations ($\omega_{\scriptscriptstyle{W}}$), H-bonds stretching vibrations ($\omega_{\scriptscriptstyle{H}}$) and nucleoside vibrations ($\omega_{\scriptscriptstyle{N}}$), respectively. From the equations (\ref{eq15})-(\ref{eq16}), the obtained modes of vibration significantly depend on the force constants and appropriate parameters.
	
	The ratio between the amplitudes of coupling vibrations may be obtained from the system of equations (\ref{eq11}) in the following form:
	\begin{equation}\label{eq17}
	\frac{\tilde{\theta}\ell}{\tilde{r}}=\frac{2k_{0}C_{1}C_{2}\frac{m}{M}}{\omega^{2}-\beta_{0}-2\alpha_{0}\sin^2\theta_{0}-k_{0}C_{1}\frac{m}{M}} .
	\end{equation}
	
	The amplitudes of vibrations have been estimated accordingly to the \cite{volkov1987conformation}. To obtain the amplitudes of the DNA with water molecules vibrations ($\tilde{\theta}$, $\tilde{\xi}$ and $\tilde{r}$), the solutions (\ref{eq10}) were substituted to the equations of potential energy (\ref{eq7}),(\ref{eq8}) and resulting expression was averaged over the time. As a result, the averaged potential energy has the form:
	\begin{equation}\label{eq18}
	\langle U\rangle=\frac{1}{8}\bigl[U^{+}+U^{-}\bigr],
	\end{equation}
	where \begin{equation}
	U^{+}= 2\alpha \ell_{a}^{2}\tilde{\theta}^{2}+\beta \tilde{\theta}^{2}+k\left(\tilde{\theta}\ell C_{1}+2\tilde{r}C_2\right)^{2} \text{and}  \ U^{-}= \beta \tilde{\xi} ^{2}+k\left(\tilde{\xi}\ell C_{1}\right)^{2}.\notag
	\end{equation}
	
	According to Boltzmann hypothesis about the uniform energy distribution by degrees of freedom the average energy per one degree of freedom is equal to $k_{\scriptscriptstyle{B}}T/2$, where $T$ is the temperature and $k_{\scriptscriptstyle{B}}$ is the Boltzmann constant \cite{anselm1981introduction}. Using the amplitudes ratio (\ref{eq17}) and equation for averaged energy (\ref{eq18}), the vibrational amplitudes of the water molecules can be determined is as follows:
	
	\begin{equation}\label{eq19}
	\tilde{r}=2\sqrt{\frac{k_{\scriptscriptstyle{B}}T}{U^{+}_{\tilde{r}}}}.
	\end{equation}
	Here $U^{+}_{\tilde{r}}$ has the form:
	\begin{equation}
	U^{+}_{\tilde{r}}=2\alpha \sin^{2}\theta_0\left(\frac{\tilde{\theta}\ell }{\tilde{r}}\right)^{2}+\frac{\beta}{\ell^2}\left(\frac{ \tilde{\theta}\ell }{\tilde{r}}\right)^{2}+k\left(\frac{\tilde{\theta}\ell }{\tilde{r}}C_{1}+2C_2\right)^{2}.\notag
	\end{equation}
	
	Using the formulae (\ref{eq15}),(\ref{eq16}),(\ref{eq18}) and (\ref{eq19}) the frequencies and  amplitudes of the vibrations may be determined.

	\section{Model parameters}\label{sec:3}
	
	To estimate the frequencies of vibrations of the system, the structural parameters ($\theta_0, \ell, \phi_0 $, $M$ and $m$) and the force constants ($\alpha, \beta, k$) of the model should be determined. The parameters $\theta_0, \ell$ for  \emph{B}-DNA double helix have been taken from \cite{volkov1987conformation,volkov1991theory}. The force constant of H-bonds stretching in the complementary base pairs  $\alpha$ and the force constant of nucleoside vibrations $\beta$ are equal to  80 \ kcal/mol{\AA}$^2$ and 40 \ kcal/mol, respectively. The values of equilibrium angle $\theta_{0}$ and reduced length $\ell$ are equal to 28$^{\circ}$ and 4.9 {\AA}, respectively \cite{volkov1987conformation,volkov1991theory}. The mass of nucleoside $M$ has been taken 199 u.m.a.. The distances $a$ and $b$ from selected centre of coordinate system (water molecule) to the suspension point of nucleosides (backbone) are approximately equal 7.3 {\AA} and 1.7 {\AA}, respectively (Fig.\ref{fig:2}b). Using the X-ray structure of the Drew-Dickerson dodecamer \cite{tereshko1999hydrat} the angle $\phi_{0}$ have been calculated. The average value of angle $\phi_{0}$ over the water-bounded pairs in the minor groove is equal to 39$^\circ$. The mass $m$ of water molecule is 18 u.m.a..

In the present work, the force constant of water vibrations $k$ has been estimated using the potential of mean force (PMF) derived from the molecular dynamics simulation data \cite{perepelytsya2018hydration}. Water molecules are trapped in the potential well and to obtain its shape, the PMF has been used. The PMFs for water molecules in each water-bonded pair have been calculated from the definition \cite{chandler1987introduction}:
\begin{equation}\label{eq20}
	U_{\scriptscriptstyle{PMF}}=-k_{\scriptscriptstyle{B}}T\ln(g(r)),
	\end{equation}
	where $k_{\scriptscriptstyle{B}}$ is the Boltzmann constant, $T$ is the temperature and $g(r)$ is the radial distribution function (RDF).
The RDFs have been derived using molecular dynamics simulations trajectories obtained in  \cite{perepelytsya2018hydration}, where the system of Drew-Dickerson dodecamer surrounded by water molecules and counterions was studied. The RDFs have been built for oxygen atoms of water molecules with the respect to reference atoms of nucleotides: O$_{2}$ of cytosine and thymine, N$_{3}$ of guanine and adenine.  The radial distribution functions have been calculated using the plugin \cite{levine2011fast} implemented to the VMD software \cite{humphrey1996vmd}. RDFs have been obtained for each base bounded with water molecule in the DNA minor groove. 
The potential of mean force calculated by (\ref{eq20}) has been fitted by the fourth degree polynomial:
	\begin{equation}\label{eq21}
	U_{\scriptscriptstyle{PMF}}\approx A+B_{1}r+B_{2}r^2+B_{3}r^3+B_{4}r^4,
	\end{equation}
	where $A,B_1, B_2, B_3,B_4$ are the fitting coefficients and $r$ is the distance.
	
	From the other side, the potential of mean force may be presented as Taylor expansion of potential energy with respect to the equilibrium point: 
	\begin{equation}\label{eq22}
	\begin{split}
	U_{\scriptscriptstyle{PMF}} \approx 	& \ U_{\scriptscriptstyle{PMF}}\vert_{r=r_0}+\frac{1}{2!}\frac {\partial^2 U_{\scriptscriptstyle{PMF}}}{\partial r^2}\vert_{r=r_0} (r-r_0)^2+\\
	+& \frac{1}{3!}\frac {\partial^3 U_{\scriptscriptstyle{PMF}}}{\partial r^3}\vert_{r=r_0}(r-r_0)^3+ 
	+\frac{1}{4!}\frac {\partial^4 U_{\scriptscriptstyle{PMF}}}{\partial r^4}\vert_{r=r_0} (r-r_0)^4, 
	\end{split}
	\end{equation}
	where $r_0$ is the position of minimum.
	
	Using equations (\ref{eq21}) and (\ref{eq22}), the force constant $k$ has been obtained as follows:
	\begin{equation}\label{eq23}
	k=\frac {\partial^2 	U_{\scriptscriptstyle{PMF}}}{\partial r^2}\vert_{r=r_0}= 2B_{2}+6B_{3}r_{0}+12B_{4}r_{0}^2.
	\end{equation}
	
	The force constants $k$ have been estimated for each nucleotide in the minor groove of the Drew-Dickerson dodecamer using the equation (\ref{eq23}). The obtained values of $k$ have been averaged for the water-bounded nucleotide pairs. The calculated force constants for water molecules with nucleic bases in water-bounded pairs in the minor groove are shown in \ref{tab:1}.
	\begin{table*}
	\caption{The values of force constant $k$ averaged for the water-bonded pairs derived from MD simulation. In the present work, the water-bounded pair is the nucleic bases that interact with the water molecules belong to different nucleotide pairs.}
	\label{tab:1}
	\begin{tabular*}{1\textwidth}{@{\extracolsep{\fill}}ccccccccc}
		\hline\noalign{\smallskip}
		Water-bounded pair&G-G& A-C &A-T&T-T &T-A&C-A&G-G\\
		\noalign{\smallskip}\hline\noalign{\smallskip}
		$k$,(kcal/mol{\AA}$^2$) & 37 &38 & 31 & 26 &32 &40 &39\\ 
		\noalign{\smallskip}\hline
	\end{tabular*}
\end{table*}
	\section{Frequencies and amplitudes of vibrations}\label{sec:4}
	
	The values of frequencies of vibrational modes have been estimated using the formulas (\ref{eq15}),(\ref{eq16}) and the parameters determined in the previous section. The obtained frequencies of vibrations are presented in the Table \ref{tab:2}.
	\begin{table*}
	\small
	\caption{ The frequency of vibrations of DNA with water molecules. $\omega_{\scriptscriptstyle{W}}$ is the frequency of water vibrations; $\omega_{\scriptscriptstyle{H}}$ is the frequency  of H-bond stretching vibrations; $\omega_{\scriptscriptstyle{N}}$ is the frequency of nucleoside vibrations.}	
	\label{tab:2}
	\begin{tabular*}{1\textwidth}{@{\extracolsep{\fill}}ccccccccc}
		\hline\noalign{\smallskip}
		Water-bounded pair&G-G&A-C&A-T&T-T&T-A&C-A&G-G\\
		\noalign{\smallskip}\hline\noalign{\smallskip}
		$\omega_{\scriptscriptstyle{W}}$,(cm$^{-1}$)  & 199 &201& 181 & 167 &184&205 &203\\
		$\omega_{\scriptscriptstyle{H}}$,(cm$^{-1}$) & 47 & 47 &  47 & 47 & 47 &47& 47 \\
		$\omega_{\scriptscriptstyle{N}}$,(cm$^{-1}$) &16 &16&15&14&15&16&16\\
		\noalign{\smallskip}\hline
	\end{tabular*}
\end{table*}
	The mode of the water molecule vibration in the DNA minor groove $\omega_{\scriptscriptstyle{W}}$ ranges from 167 to 205 cm$^{-1}$. Our calculations have shown that the frequency of water vibration depends on nucleotide sequence: the frequency $\omega_{\scriptscriptstyle{W}}$ increases from the centre of the Drew-Dickerson fragment (CGCGAATTCGCG) to its periphery where the cytosine and guanine bases are located. The lowest value of the frequency of water molecule vibration (167 cm$^{-1}$) is observed in the centre of the spine of hydration. The analysis has shown that the water dynamics in the water-bounded pairs may depend on type of acceptor atoms of nucleic bases (binding site).  
	
	The frequency of the mode of H-bond stretching vibrations ($\omega_{\scriptscriptstyle{H}}$) is about 47 cm$^{-1}$. It is about twice lower than in the case of the approach of the four-mass model and experimental data (about 85 cm$^{-1}$)  \cite{urabe1985collective,volkov1987conformation,volkov1991theory}. According to the four-mass model, the H-bond stretching occurs due to the symmetrical rotations of nucleosides around phosphates and motions of the masses of phosphates groups. However, in the present model, the phosphates are fixed and in this case, the H-bonds stretching vibrations in complementary pairs are caused only by pendulum-nucleosides vibration around the phosphate groups. At the same time, the analysis, performed in the present work, showed that variation of the force constant $\alpha$ does not influence the water molecule vibrations.  
	
	The frequency of the mode $\omega_{\scriptscriptstyle{N}}$ ranges from 14 to 16 cm$^{-1}$ which depends on nucleoside type. The calculated result in good agreement with results of four-mass model \cite{volkov1987conformation,volkov1991theory}, where this mode is referred to pendulums-nucleoside vibrations. In turn, calculated frequencies of pendulums-nucleoside vibrations ($\omega_{\scriptscriptstyle{N}}$) are in good agreement with experimental studies. In this spectra range, experimental data\cite{urabe1981low,urabe1991low,woods2006effect} have defined mode around 20 cm$^{-1}$, which value significantly depend on macromolecule hydration and associated it with vibration of staked bases.
	
	To determine a character of the DNA conformational vibration with water molecules in the hydration spine of the DNA minor groove, the amplitudes of the displacements of the nucleosides and water in water-bouned pairs are calculated using formulae (\ref{eq22}). The amplitudes of the vibrations describe the displacements of the mass centres from their equilibrium positions. The obtained amplitudes of water and nucleosides masses displacements are presented in Table \ref{tab:3}. The value of H-bonds stretching in complementary pairs have been also calculated using expression $\delta = l_a\tilde{\theta}$. 
	
	As follows from the equations (\ref{eq11}) and (\ref{eq12}), the amplitudes of vibration $\tilde{\theta}$ and $\tilde{\xi}$ characterize symmetrical and asymmetrical vibrations of pendulum-nucleosides, respectively, while the amplitudes of vibration $\tilde{r}$ characterize vibrations of the water molecule in water-bounded pairs. Analysis of obtained amplitude values shows that asymmetrical movements of nucleosides occur without participation of water molecule and H-bonds stretching vibrations. Whereas symmetrical movements of the masses in the model occur with H-bonds stretching in the base pairs and water molecules vibration.
	\begin{table} 
	\small
	\caption{The amplitudes of the DNA conformational vibrations with the water molecules which are located in the minor groove of double helix. $\tilde{\theta}$ and $\tilde{\xi}$ are the amplitudes of pendulum-nucleoside vibrations; $\tilde{r}$ is the amplitude of water molecule vibrations in water-bonded pairs; $\delta$ is the value of H-bonds stretching in base pairs. $\omega_{\scriptscriptstyle{W}}$,$\omega_{\scriptscriptstyle{H}}$,$\omega_{\scriptscriptstyle{N}}$ are the frequency of vibration (cm$^{-1}$).}
	\label{tab:3}
	\begin{tabular*}{1\textwidth}{@{\extracolsep{\fill}}ccccccccc}
		& & &&\textbf{ G-G }&&&& \\
		\hline\noalign{\smallskip}\small 
		&&$\tilde{\theta}$ ($^{\circ}$) & &$\tilde{r}$(pm) & &	$\tilde{\xi}$ ($^{\circ}$) &&$\delta$ (pm)  \\
		\hline\noalign{\smallskip}
		$\omega_{\scriptscriptstyle{W}}$  & & 0.045& &14.0& &0&& 0.18\\
		$\omega_{\scriptscriptstyle{H}}$ && -2.97 && 3.85 & & 0&& -11.9\\
		$\omega_{\scriptscriptstyle{N}}$ &&0& &0&&8.9&& 0\\
	
		\hline\noalign{\smallskip}
		& & && \textbf{A-C} &&&& \\
		\hline\noalign{\smallskip}
		$\omega_{\scriptscriptstyle{W}}$  & &0.044& &13.9& &0&& 0.18\\
		$\omega_{\scriptscriptstyle{H}}$ && -2.97 && 3.84 & & 0 && -11.9\\
		$\omega_{\scriptscriptstyle{N}}$ &&0& &0&&8.8&&0 \\
		
		\hline\noalign{\smallskip}

		& & && \textbf{A-T} &&&& \\
		\hline\noalign{\smallskip}
		$\omega_{\scriptscriptstyle{W}}$ & &0.050& &15.4& &0&&0.20 \\
		$\omega_{\scriptscriptstyle{H}}$&& -2.97 &&3.89& & 0&&-11.9\\
		$\omega_{\scriptscriptstyle{N}}$ &&0& &0&&9.4&&0 \\

		\hline\noalign{\smallskip}
		& & &&\textbf {T-T} &&&& \\
		\hline\noalign{\smallskip}
		$\omega_{\scriptscriptstyle{W}}$  & &0.055& &16.7& &0&&0.22\\
		$\omega_{\scriptscriptstyle{H}}$ && -2.97 &&3.94& &0&&-11.9\\
		$\omega_{\scriptscriptstyle{N}}$&&0& &0&&9.8&&0 \\

		\hline\noalign{\smallskip}
		& & && \textbf{T-A} &&&& \\
		\hline\noalign{\smallskip}
		$\omega_{\scriptscriptstyle{W}}$ && 0.049& &15.2& &0&&0.19 \\
		$\omega_{\scriptscriptstyle{H}}$ && -2.97 &&3.88& &0&&-11.9\\
		$\omega_{\scriptscriptstyle{N}}$ &&0& &0&&9.3&&0 \\

		\hline\noalign{\smallskip}
		& & && \textbf{C-A} &&&& \\
		\hline\noalign{\smallskip}
		$\omega_{\scriptscriptstyle{W}}$  & &0.041& &13.6& &0&&0.17 \\
		$\omega_{\scriptscriptstyle{H}}$ && -2.97&&3.83& &0&&-11.9\\
		$\omega_{\scriptscriptstyle{N}}$ &&0& &0&&8.7&& 0\\

		\hline\noalign{\smallskip}
		& & && \textbf{G-G} &&&& \\
		\hline\noalign{\smallskip}
		$\omega_{\scriptscriptstyle{W}}$  & &0.044& &13.7& &0&&0.18\\
		$\omega_{\scriptscriptstyle{H}}$ && -2.97 &&3.84& &0&&-11.9\\
		$\omega_{\scriptscriptstyle{N}}$ &&0& &0&&8.8&& 0\\
		\hline\noalign{\smallskip}
	\end{tabular*}
\end{table}

	The amplitudes of water molecules in the case of the modes $\omega_{\scriptscriptstyle{W}}$ depend on nucleotide vibration. In the centre of the spine of hydration, the amplitudes $\tilde{r}$ are slightly higher than in the ends, as well as in the case of amplitude $\tilde{\theta}$. In turn, the amplitudes of the symmetrical vibrations of pendulums-nucleosides in the case of the modes $\omega_{\scriptscriptstyle{H}}$ are similar for all selected nucleosides that are caused by constant value of vibrational mode of H-bonds stretching for all selected pairs (Table \ref{tab:2}). From Table \ref{tab:3} follows that vibrational modes $\omega_{\scriptscriptstyle{N}}$ caused by asymmetrical vibrations of pendulums-nucleosides are not affect on the amplitudes of vibrations of water molecules and H-bonds stretching.
	
The analysis of calculated amplitudes of displacements demonstrates that the water molecules vibrations are interrelated with the conformational vibrations of the DNA double-helix. The obtained frequencies and amplitudes of vibrations show that dynamics of water molecule depends on nucleotide sequence and binding sites. 
	\begin{figure}[H]
	\includegraphics[width=0.75\textwidth]{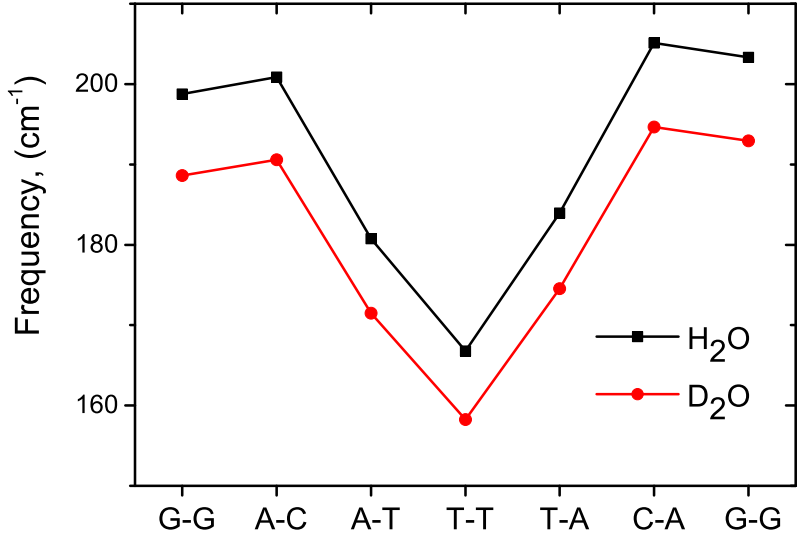}
	\caption{\small{The frequency of vibration of light and heavy water bridging bases in water-bounded pairs in the DNA minor groove. The black line with squares identifies the modes of vibration of light water (H$_2$O) and red line with circles identifies the modes of vibration of heavy water (D$_2$O).}}
	\label{fig:3}
\end{figure}
	
	\section{Theory and experiment}\label{sec:5}		
	The low-frequency vibrational modes calculated in the present work should be observed in the experimental spectra of DNA. At the same time, the observation of these modes may be a complicated task due to the presence of the translational water vibrations,
	observed around 180 cm$^{-1}$ \cite{walrafen1964raman}. The Raman spectroscopy experiments \cite{galvin2011extreme} for liquid water showed that symmetric and asymmetric stretch of H-bonds are about 111 and 157 cm$^{-1}$, respectively, while coupled mode of translation and rotation vibrations is about 220 cm$^{-1}$. Therefore, our calculations should be compared with experimental data observed in the frequency range from 150--220 cm$^{-1}$, where our calculated frequencies are obtained. 
	
	In the far-infrared spectroscopic experiments \cite{powell1987investigation} for various polynucleotides at 300 K, the DNA vibrations modes lower than 250 cm$^{-1}$ are present. In particular, the characteristic peaks in the spectra of poly(dA)poly(dT) around 170 and 214 cm$^{-1}$ and in the spectra of polynucleotide poly(dA-dT)poly(dA-dT) around 195 cm$^{-1}$ are indicated. Our results show that vibrational modes of water molecules bridging bases in nucleotide sequence AATT are around 181,167 and 184 cm$^{-1}$ (Table \ref{tab:2}). The vibrational modes of poly(dG)poly(dC) indicated by experimental data ranged from 160 to 235 cm$^{-1}$, while our calculations show vibrational modes within the range 199--205 cm$^{-1}$. The obtained results sufficiently agree with experimental data. Therefore, it may be expected that the modes observed by spectroscopy experiments may characterize the modes of water molecules vibrations.
	
	To find the mode of water vibrations in the hydration spine of the DNA minor groove the dynamics of the double helix with heavy water (D$_2$O) may be considered. For this analysis, the developed approach was used. The frequencies of vibrations of heavy water in the DNA minor groove have been obtained using equations (\ref{eq15}) and (\ref{eq16}). Our calculations have shown that the frequency mode of heavy water vibrations ranges from 158 to 195 cm$^{-1}$. The comparison of the frequencies of vibrations of heavy water and light water in the DNA minor groove are shown in Fig. \ref{fig:3}. The obtained frequencies modes of vibrations of heavy water softened for about 10 cm$^{-1}$ comparing to H$_2$O that may be explained due to the increasing mass. Decreasing in the frequency value should be observed in the experimental vibrational spectra. In the case of liquid water, the experimental data for D$_2$O has shown that the mode of translation vibrations soften for about 10 cm$^{-1}$ compared to the H$_2$O \cite{draegert1966far}. Therefore, the comparison of the low-frequency spectra DNA with heavy water can provide additional information for identifying the water modes in the DNA minor groove.
	
	\section{Conclusion}
	The model of the low-frequency water vibrations in the DNA minor groove has been developed. Our calculations have shown that the frequency of translational vibrations of water molecules within the frequency range 167--205 cm$^{-1}$. The obtained values of the frequencies and amplitudes of vibrations indicate that the character of the dynamics of water molecules significantly depends on the conformational vibrations of the DNA double helix and the binding site of water molecules with atoms of nucleotide bases. The calculated modes of water molecules vibrations observed in the same spectra range as translational vibrations of the water in the bulk phase. To distinguish the vibrations of water molecules in the DNA minor groove from those in the bulk, the dynamics of DNA with heavy water has been considered. The calculations have shown that the frequency of vibration of heavy water in the DNA minor groove ranges from 158 to 195 cm$^{-1}$. Decreasing of vibrational mode of heavy water for about 10 cm$^{-1}$ compared to the light water was obtained in our calculations and is expected to be observed in the experimental spectra. The mode softening due to the heavy water may be used for identifying the mode of water vibration in the spine of hydration in DNA minor groove.
	
	\section{Acnowledgements}\label{sec:6}
	The authors gratefully acknowledge Prof. Sergey Volkov and the colleagues
	of the Laboratory of Biophysics of Macromolecules of the BITP for the
	stimulative discussion. The present work was partially supported by the
	Project of the National Academy of Sciences of Ukraine (0120U100855).

\end{document}